# Non-Monotonic Size-Dependent Exciton Radiative Lifetime in CsPbBr$_3$ Nanocrystals

*Abdullah S. Abbas[1,†], Daniel Chabeda[2], Daniel Weinberg[2,3], David T. Limmer[2,3,7], Eran Rabani[2,3,8], A. Paul Alivisatos[1,2,3,4,5,6,§,*]*

[1]Department of Materials Science and Engineering, University of California, Berkeley, Berkeley, California 94720, United States

[2]Department of Chemistry, University of California, Berkeley, Berkeley, California 94720, United States

[3]Materials Sciences Division, Lawrence Berkeley National Laboratory, Berkeley, California 94720, United States

[4]Department of Chemistry, The University of Chicago, Chicago, Illinois 60637, United States

[5]Pritzker School of Molecular Engineering, The University of Chicago, Chicago, Illinois 60637, United States

[6]James Franck Institute, The University of Chicago, Chicago, Illinois 60637, United States

[7]Kavli Energy NanoScience Institute, Berkeley, California 94720, United States

[8]The Raymond and Beverly Sackler Center of Computational Molecular and Materials Science, Tel Aviv University, Tel Aviv 69978, Israel

Present addresses:

[†]Department of Chemistry, The University of Chicago, Chicago, Illinois 60637, United States

[§]Department of Chemistry and Pritzker School of Molecular Engineering, The University of Chicago, Chicago, Illinois 60637, United States

[*]To whom correspondence may be addressed: paul.alivisatos@uchicago.edu

ABSTRACT

Lead-halide perovskite nanocrystals have recently emerged as desirable optical materials for applications such as coherent quantum light emitters and solid-state laser cooling due to their short radiative lifetime and near-unity photoluminescence quantum yield. Here, we investigate the effect of CsPbBr$_3$ nanocrystal size on the radiative lifetime under ambient conditions. High-quality nanocrystals, with monoexponential time-resolved photoluminescence decay behaviors, unveil a non-monotonic trend in radiative lifetime. This non-monotonicity appears to reflect a behavior common among II-VI (CdSe) and perovskites semiconducting nanocrystals. We find that large nanocrystals in the weak quantum confinement regime exhibit long radiative lifetimes due to a thermally accessible population of dim states. Small nanocrystals within the strong quantum confinement regime, surprisingly, also show long radiative lifetimes, due however to a substantial reduction in oscillator strength. Nanocrystals in the intermediate quantum confinement regime displays the shortest radiative lifetime, as their oscillator strength is enhanced relative to particles in the strong confinement regime, but do not have sufficient low-lying dim states like the large particles to counteract this affect. These findings shed light on the impact of nanocrystal size on radiative lifetime and pave the way for tailored optical materials in various optical applications.

Colloidal nanocrystals, also known as quantum dots (QDs), have attracted considerable attention due to their straightforward and inexpensive solution-phase synthesis and their wide-range spectral tunability from the quantum confinement effect[1,2]. Lead-halide perovskite nanocrystals, in particular cesium lead bromide ($CsPbBr_3$), have emerged in the past few years as highly desirable and efficient optical materials due to their facile synthesis, narrow emission linewidth, and near-unity photoluminescence quantum yield (PLQY) owing to their low trap density and tolerance for shallow surface defects[3–12]. The photoluminescence quantum yield is in part dictated by the radiative lifetime, or the time between generation and radiative recombination of an exciton. Nanocrystals with near-unity PLQY and short radiative lifetimes have possible applications in coherent quantum light emitters[13] and solid-state laser cooling[14]. For example, to observe laser cooling in solid-state materials, two processes must occur in rapid succession: exciton up-conversion followed by radiative recombination, resulting in anti-Stokes photoluminescence[14]. A shorter radiative lifetime enhances the likelihood of avoiding non-radiative pathways that induce heating, thereby maximizing the cooling probability. One way to tune the radiative lifetime is through the nanocrystal size[15,16]. For nanocrystals, three distinct quantum confinement regimes dictated by the Bohr radius – weak, intermediate, and strong – are commonly observed. Each regime exhibits unique electronic structure and optical properties, influencing factors such as the radiative lifetime. Notably, $CsPbBr_3$ nanocrystals have a relatively large Bohr diameter (~7nm), enabling experimental exploration across all three confinement regimes[3].

In this work, we explore the effect of nanocrystal size on radiative lifetime under ambient conditions. High-quality $CsPbBr_3$ nanocrystals were synthesized to ensure monoexponential decay behaviors, enabling direct extraction of the radiative lifetime. Unexpectedly, we observe a non-monotonic trend in the radiative lifetime. The smaller nanocrystals within the strong confinement regime and larger nanocrystals within the weak confinement regime surprisingly exhibited long radiative lifetimes, while the intermediate confinement regime displays the shortest radiative lifetime. In previous work investigating size-dependent lifetime in $CsPbBr_3$ under low temperature conditions (~4-5 K), revealed an exponential trend, with radiative lifetime decreasing as the size increases[13,15]. Under ambient conditions, however, the work of Krieg *et al* shows the reverse trend[16]. It was rationalized that higher lying states become thermally accessible at ambient conditions and introduce a delay in the radiative recombination of excitons[16]. Neither of these studies thoroughly explored the inflection point marking the transition from the intermediate to the strong confinement regime.

We observed a symmetry lowering transition from the cubic to the orthorhombic phase[17–21] of the $CsPbBr_3$ nanocrystal system in a coincidentally similar size regime, aligning with the inflection point observed in the radiative lifetimes. Both X-ray diffraction (XRD) patterns and molecular dynamics simulations conducted on $CsPbBr_3$ nanocrystals with varying sizes confirm the occurrence of this phase transition. However, we find no evidence that this phase transition has a direct influence on the radiative lifetime. Rather, in larger nanocrystals within the weak confinement regime, we ascribe the lengthening of the radiative lifetime to a population of high-lying dim states that are thermally accessible at room temperature. These states effectively decrease the population of the lowest bright state under ambient conditions, resulting in a delay in the radiative lifetime for weakly confined nanocrystals where the density of states is high. This observation finds support from a previous work where the lengthening of radiative lifetime is attributed to the mixing of S state with higher P state that are symmetry forbidden[16]. Analysis of the present result in comparison to Krieg et al is presented in the Supporting Information Fig. S7. The smaller nanocrystals within the strong confinement regime, show longer radiative lifetimes due to a reduction in oscillator strength. In contrast, intermediately confined nanocrystals exhibit the shortest radiative lifetime, as the effect of states mixing is reduced in this regime along with an enhancement in the oscillator strength. The non-monotonic trend in radiative lifetime observed in the present study appears to reflect a behavior common among intensively studied II-VI (CdSe) [22–25] and perovskite semiconducting nanocrystals. In all, this study provides a thorough insight into the size-dependent radiative lifetime in $CsPbBr_3$ nanocrystals, paving the way to engineer materials with tailored decay rates to meet diverse optical application needs.

**RESULTS & DISCUSSION**

We investigated the effect of nanocrystal size on optical properties, focusing on the radiative lifetime. Over 30 high-quality nanocrystals samples were synthesized following a modified version of previous reports (details in the Method section) to minimize non-radiative pathways and to directly extract the radiative lifetime. Fig. 1a shows absorption and photoluminescence spectra of $CsPbBr_3$ nanocrystals in the strong, intermediate, and weak confinement regime. Fig. 1c, d, and e show their corresponding Transmission Electron Microscopy (TEM) images (larger area shown in Fig. S1) along with calculated size distributions[26], illustrating highly monodisperse ensemble. Photoluminescence Full-Width at Half Maximum (FWHM) and Stokes shift are extracted for different nanocrystal sizes revealing a linear trend. Both the FWHM and Stokes shift exhibit an increase as nanocrystal sizes decrease, which has been observed previously[27,28]. This linear trend is correlated between the size of the nanocrystals and the strength of coupling to high-energy optical phonons (2-17 meV)[28,29]. Specifically, smaller nanocrystals exhibit stronger coupling to these phonons due to enhanced charge density and subsequent polaron formation[30]. This coupling results in large fluctuations in the bandgap[31], consequentially causing broader emission linewidths and larger Stokes shift as evident in Fig. 1a and b.

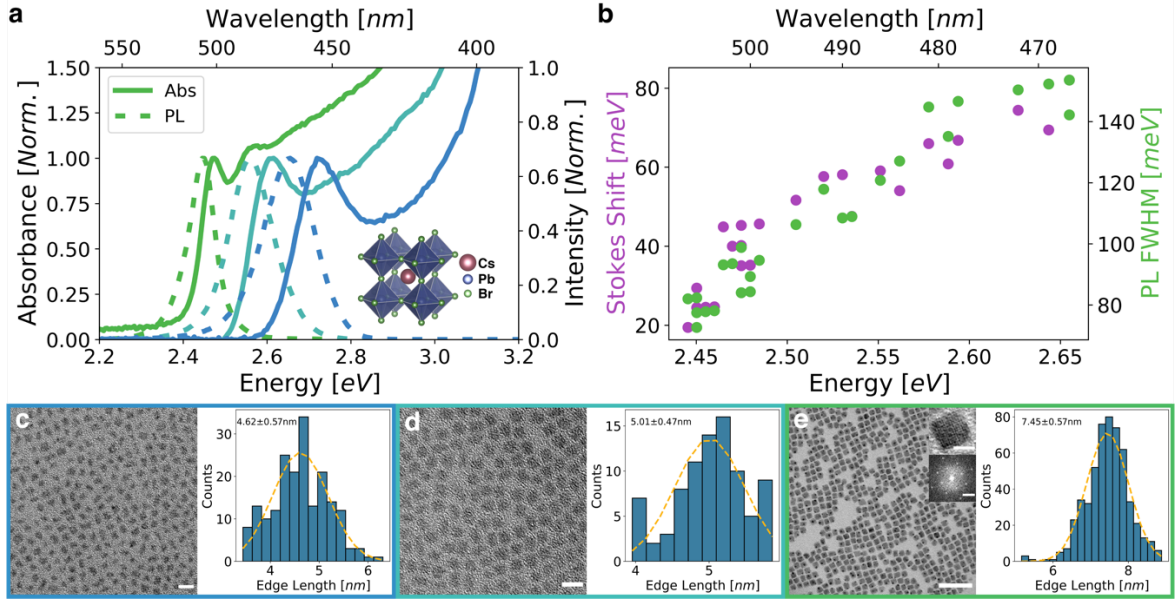

**Figure 1: Optical and structural characterizations of CsPbBr$_3$ nanocrystals with varying sizes. a**, steady-state absorption, and photoluminescence (PL) spectra of selected nanocrystals in weak, intermediate, and strong quantum confinement regime, with inset showing the crystal structure created with Vesta[32]. **b**, Stokes shift and full width at half max as a function of bandgap energy are plotted displaying a linearly increasing trend as nanocrystal sizes decrease. **c**, **d**, **e** are transmission electron microscope (TEM) images along with their corresponding size distributions fitted with gaussian curve to extract the average nanocrystal size. TEM images scale bars are as follow: **c,** 10 nm, **d**, 10 nm, and **e**, 50 nm with inset showing high resolution TEM of a single nanocrystal with 5 nm scale bar and its corresponding Fast Fourier Transform (FFT) with 2 nm$^{-1}$ scale bar.

Each synthesized distinct nanocrystal sample underwent characterization to determine photoluminescence quantum yield (PLQY) and time-resolved photoluminescence (TRPL) to quantify the radiative lifetime ($\tau_r$), which is inversely related to the radiative rate ($\kappa_r$), using the following relation:

$$PLQY = \frac{\frac{1}{\tau_r}}{\frac{1}{\tau_r} + \frac{1}{\tau_{nr}}} = \frac{\frac{1}{\tau_r}}{\frac{1}{\langle\tau\rangle}} \qquad (1)$$

$$\tau_r = \frac{\langle\tau\rangle}{PLQY} \qquad (2)$$

$$\kappa_r = \frac{1}{\tau_r} \qquad (3)$$

where $\langle\tau\rangle$ is total lifetime extracted by deconvoluting with the instrument response function (IRF) and subsequentially fitting the initial two decades of TRPL decays, as shown in Fig. 2, using a monoexponential function as follow:

$$I(t) = \int_{\infty}^{t} IRF(t') * A\, e^{-\frac{t-t'}{\tau}}\, dt' \qquad (4)$$

The tabulated data in Fig. 2 shows the near-unity PLQY CsPbBr$_3$ nanocrystals across the three confinement regimes. Notably, the strongly and weakly confined nanocrystals exhibit longer radiative lifetime (or slower radiative rate) compared to the intermediately confined nanocrystals.

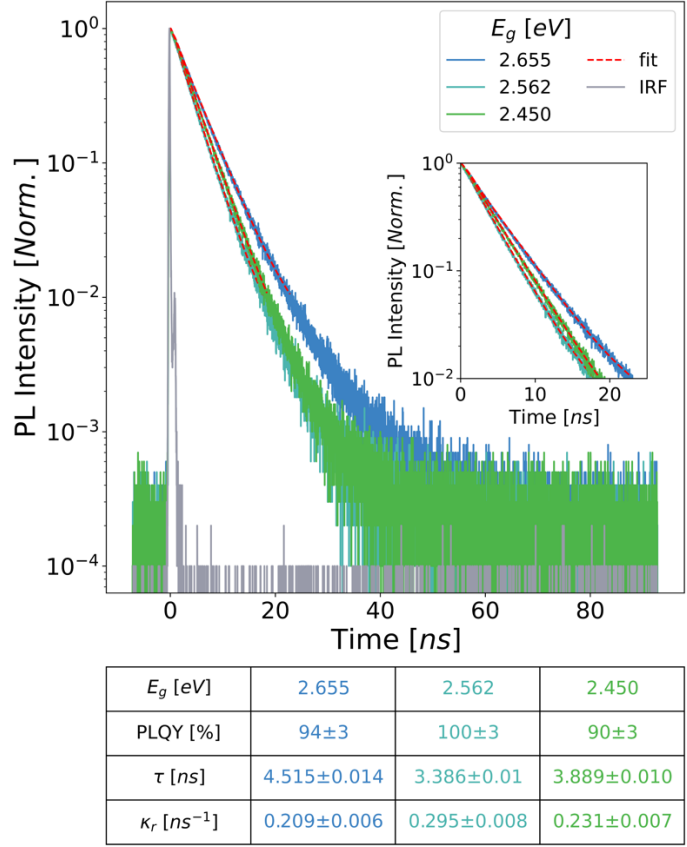

| $E_g$ [eV] | 2.655 | 2.562 | 2.450 |
| --- | --- | --- | --- |
| PLQY [%] | 94±3 | 100±3 | 90±3 |
| $\tau$ [ns] | 4.515±0.014 | 3.386±0.01 | 3.889±0.010 |
| $\kappa_r$ [ns$^{-1}$] | 0.209±0.006 | 0.295±0.008 | 0.231±0.007 |

**Figure 2: Time-resolved photoluminescence (TRPL) characterization.** Three selected nanocrystals sizes in strong, intermediate, and strong quantum confinement regimes are assessed using TRPL. Decays are deconvoluted with instrument response function (IRF) and the initial two decades are fitted to a monoexponential function to extract the total lifetime (inset). Absolute photoluminescence quantum yield (PLQY) was measured to calculate the radiative rates for each nanocrystals. The resulting data, including bandgap ($E_g$), PLQY, total lifetime, and radiative rate, are tabulated, revealing a non-monotonic trend that emerges in lifetimes and radiative rates.

We conducted further measurements of PLQY and TRPL across different nanocrystal sizes within each of the three confinement regimes under ambient conditions. Surprisingly, our findings reveal a non-monotonic trend in both radiative lifetime and radiative rate, as shown in Fig. 3a and b. We find, in confirmation of earlier reports[16], that the weakly confined nanocrystals exhibit longer radiative lifetimes. However, our measurements on strongly confined nanocrystals unexpectedly display similar long radiative lifetimes. Interestingly, we observe the shortest radiative lifetimes within the intermediate confinement regime. Thus, there exists some optimal nanocrystal size for applications where fast radiative recombination is desirable.

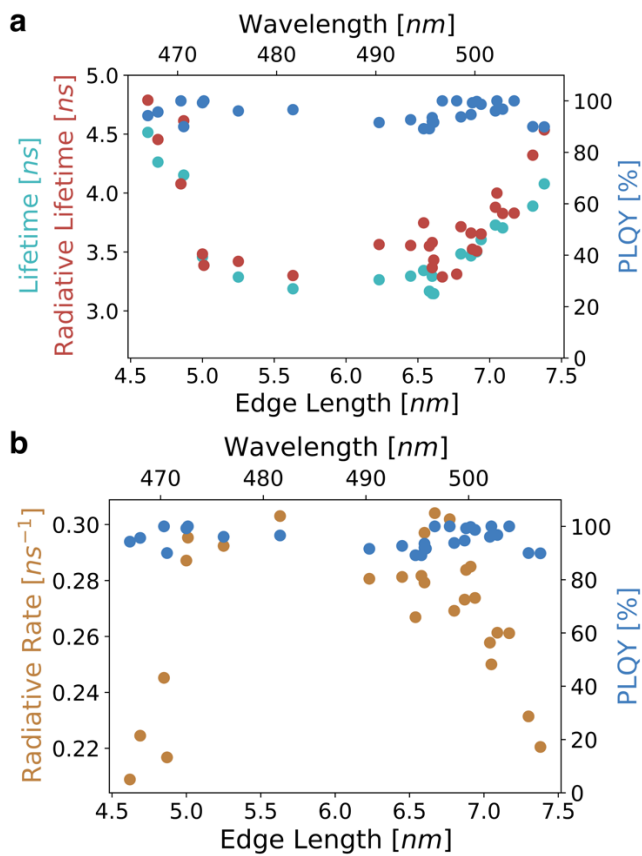

**Figure 3: Size-dependent exciton radiative lifetimes and rates of CsPbBr$_3$ nanocrystals. a**, Measured total lifetime under ambient conditions along with calculated radiative lifetimes via the absolute photoluminescence quantum yield (PLQY) as a function of nanocrystal size exhibit an unexpected non-monotonic trend. Larger nanocrystals in weak confinement regime show lengthened radiative lifetimes that decrease as the nanocrystal size transition into the intermediate confinement regime. Surprisingly, the trend reverses with decreasing nanocrystals size reaching

the strong confinement regime. **b**, radiative rates, the inverse of radiative lifetimes, showcasing an optimal size associated with maximum radiative rate.

To understand the underlying factors that may contribute to the lengthened radiative lifetime, we delved into a comprehensive examination of the crystal structure. In Fig. 4a, the powder X-ray diffraction (XRD) patterns of the nanocrystals with sizes spanning the three confinement regimes are illustrated. The nanocrystals in the strong confinement regime exhibits a cubic phase. As the size of the nanocrystals increases, a transition into an orthorhombic phase is observed[17–21]. This phase transition, as discussed below, is further corroborated by molecular dynamics simulations at 300 K, as shown in Fig 4b.

Molecular dynamics simulations as implemented in LAMMPS were performed on $CsPbBr_3$ nanocubes with edge lengths of 2 – 6 nm using a previously parametrized force field[33]. The average structural symmetry at 300 K was computed over nanosecond simulation lengths to resolve the size-dependent phase behavior. To produce a quantitative metric of the structural symmetry, we define a local order parameter, $\xi$, that distinguishes the common cubic and orthorhombic phases of $CsPbBr_3$ nanocrystals. The order parameter is defined on the probability distribution of nearest neighbor halide-halide-halide (X-X-X) angles of the nanocrystal configuration,

$$\xi_i(\theta) = \frac{[P_i(\theta) - P°_{cubic}(\theta)]^2}{[P°_{ortho}(\theta) - P°_{cubic}(\theta)]^2} \qquad (5)$$

where $P_i(\theta)$ is the probability of finding an X-X-X angle of $\theta$ in the $i$-th trajectory step, and $P°(\theta)$ is the probability of observing angle $\theta$ in the perfectly symmetric structure. If the configuration along the trajectory possesses cubic symmetry, then the order parameter becomes zero. If the configuration is orthorhombic, the ratio of terms is unity, and the order parameter is one. The averaged order parameter for a certain size, $\langle \xi(\theta) \rangle$, is estimated over the simulation trajectory, and we compare it against perfectly cubic and orthorhombic nanocubes of the same size.

The probability distribution of X-X-X angles distinguishes strongly between cubic and orthorhombic CsPbBr$_3$ at two angles, 60° and 160°. For perfectly cubic perovskites, 1/6 of the angles are 60° while zero angles are 60° for perfectly orthorhombic structures. This is reversed at 160°, where perfectly cubic structures have no probability and orthorhombic structures have a 1/6 probability. Fig. S4 shows histograms illustrating the probabilities, allowing for a clear distinction of certain angles that specifically emerge in one phase but are absent in the other. We develop a composite measure of the size-dependent phase behavior of perovskite nanocrystals in our simulations by averaging $\xi(60°)$ and $\xi(160°)$ over a narrow angular bin:

$$\xi = \frac{\xi(60°) + \xi(160°)}{2}. \tag{6}$$

As evident in Fig. 4b, the smaller nanocrystals, characterized by strong quantum confinement, display a cubic phase. As the nanocrystals size increases, a transition occurs, with the phase progressively shifting towards orthorhombic symmetry – results that closely align with the experimental XRD patterns shown in Fig. 4a. simulations were also performed on analogous CsPbI$_3$ perovskite nanocrystals and a similar phase behavior observed. Comparative analysis of the size-dependent anisotropy for CsPbBr$_3$ and CsPbI$_3$ nanocrystals are presented in Fig. S5.

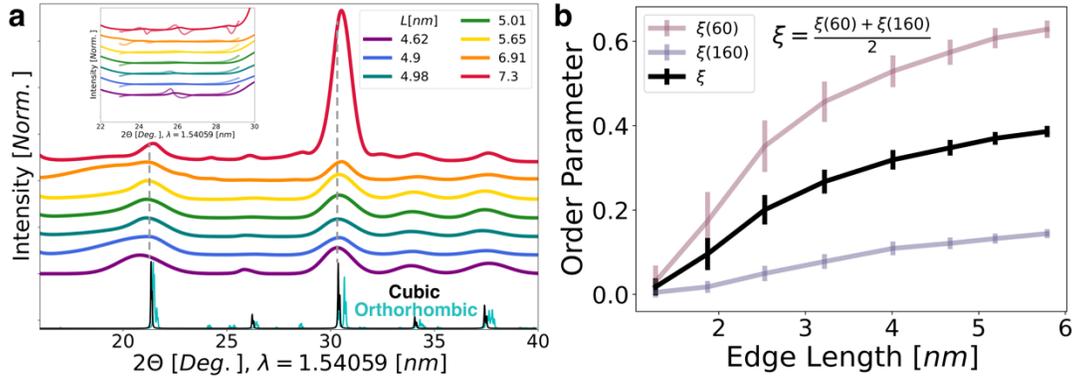

**Figure 4: X-ray diffraction (XRD) and molecular dynamics (MD) of varying CsPbBr$_3$ nanocrystal sizes. a**, XRD patterns illustrating the transition from cubic to orthorhombic symmetry. **b**, a composite order parameter that assesses the degree of crystal phase computed from MD simulation trajectories under ambient conditions in reference to a perfect cubic and orthorhombic symmetry, where 0 indicates a cubic phase and 1 means an orthorhombic phase. Error bars denote the variance of the order parameter due to fluctuations in the geometry over the simulation. The order parameter results align with XRD patterns, confirming the size-dependent phase transition from being cubic in small nanocrystals to orthorhombic in larger nanocrystals.

Further theoretical understanding of the interplay between perovskite structural and optical properties was gained by performing electronic structure calculations on the relaxed structures for varying nanocrystal sizes. Accurate excited states including electron-hole correlation and spin-orbit coupling effects were computed within the semi-empirical pseudopotential framework as described in previous reports[34]. We obtain a validated pseudopotential from Weinberg et al. describing local, nonlocal and spin-orbit coupling effects for CsPbI$_3$[34]. The electronic properties of CsPbI$_3$ and CsPbBr$_3$ show analogous trends[35]; thus, we compute radiative lifetimes using the validated CsPbI$_3$ model as a proxy to understand the measured CsPbBr$_3$ radiative lifetimes. Due to computational constraints, the investigation of nanocrystal sizes spans strongly and intermediately confined nanocrystals with sizes ranging from 3.1 – 5.7 nm. In addition, we relax nanocrystal structures at 0 K using a conjugate gradient approach to obtain static input geometries. It was verified that the 0 K structures show the same size dependent phase behavior as the 300 K simulations (see Fig. S4). This gives us three structures to compare for each size: cubic, orthorhombic, and relaxed.

We calculated the thermally averaged radiative lifetime in the time-dependent perturbation theory framework as:

$$\langle \tau_r \rangle = \frac{1}{\langle \kappa_r \rangle} = \left[ \frac{1}{Z} \sum_n e^{-\beta E_n} \frac{\omega_n^3 |\mu_n|^2}{3\pi \varepsilon_o \hbar c^3} \right]^{-1} \quad (8)$$

where $\varepsilon_o, \hbar, c$ are the vacuum permittivity, reduced Planck's constant, and speed of light, respectively. The parameters $\omega_n$ and $\mu_n$ correspond to the transition frequency and the transition dipole strength of exciton $n$. Reported radiative lifetimes are the Boltzmann thermal average over excitons. Emission from dark states occurs on a much longer timescale than from bright states, generating long- (ms) and short-time (ns) contributions to the average lifetime. We assume that the measured lifetimes are the short-time component coming from bright states in the exciton manifold. Thus, we report the Boltzmann averaged lifetime computed from states with ~ns lifetimes and ignore the ~ms contributions. The partition function is summed over all states indiscriminately. This procedure produces radiative lifetimes that agree well with experiments. A reduction in radiative lifetime is observed as nanocrystal size increases from the strongly confined to intermediately confined regime (Fig. 5a). We rationalize this decrease in radiative lifetime by analyzing the size-dependent oscillator strength of thermally accessible states of the lowest bright exciton. We find a drastic increase in the lowest bright state's oscillator strength as the nanocrystal size increases, independently of crystal symmetry (Fig. 5b). While one might expect a further decrease in radiative lifetime due to the increase in the oscillator strength, an interesting lengthening in radiative lifetime occurs. Due to the increasing density of states in larger nanocrystals, high-lying states exhibiting weak oscillator strength (i.e., dim states) become thermally accessible and lead to a decreased Boltzmann population of the bright states. The thermal population of high-lying states leads to an extension of radiative lifetimes in larger nanocrystals that results in non-monotonicity in the radiative lifetime as observed for the largest calculated nanocrystals, shown in Fig. 5a. To examine how temperature influences of non-monotonic trend, we computed the radiative lifetimes at lower temperatures (Fig. S6). Our analysis reveals a monotonic decrease as nanocrystals size decrease at lower temperatures, indicating that the dim states become thermally inaccessible. This monotonically decreasing trend aligns closely with previously reported experimental findings[13,15]. It is noteworthy to point that the competing effects of weakened oscillator strength and occupation of high-lying states occur distinctly for the strong and weak confinement regimes, respectively, leaving the intermediate confinement regime to exhibit the shortest radiative lifetime. This aligns well with the experimentally observed non-monotonic trend in size-dependent radiative lifetime in $CsPbBr_3$ nanocrystals, shown in Fig. 3. Additionally, this rationalization explains the trend we predict to be common among prototypical II-VI (CdSe)[22–25] and perovskite semiconducting nanocrystals.

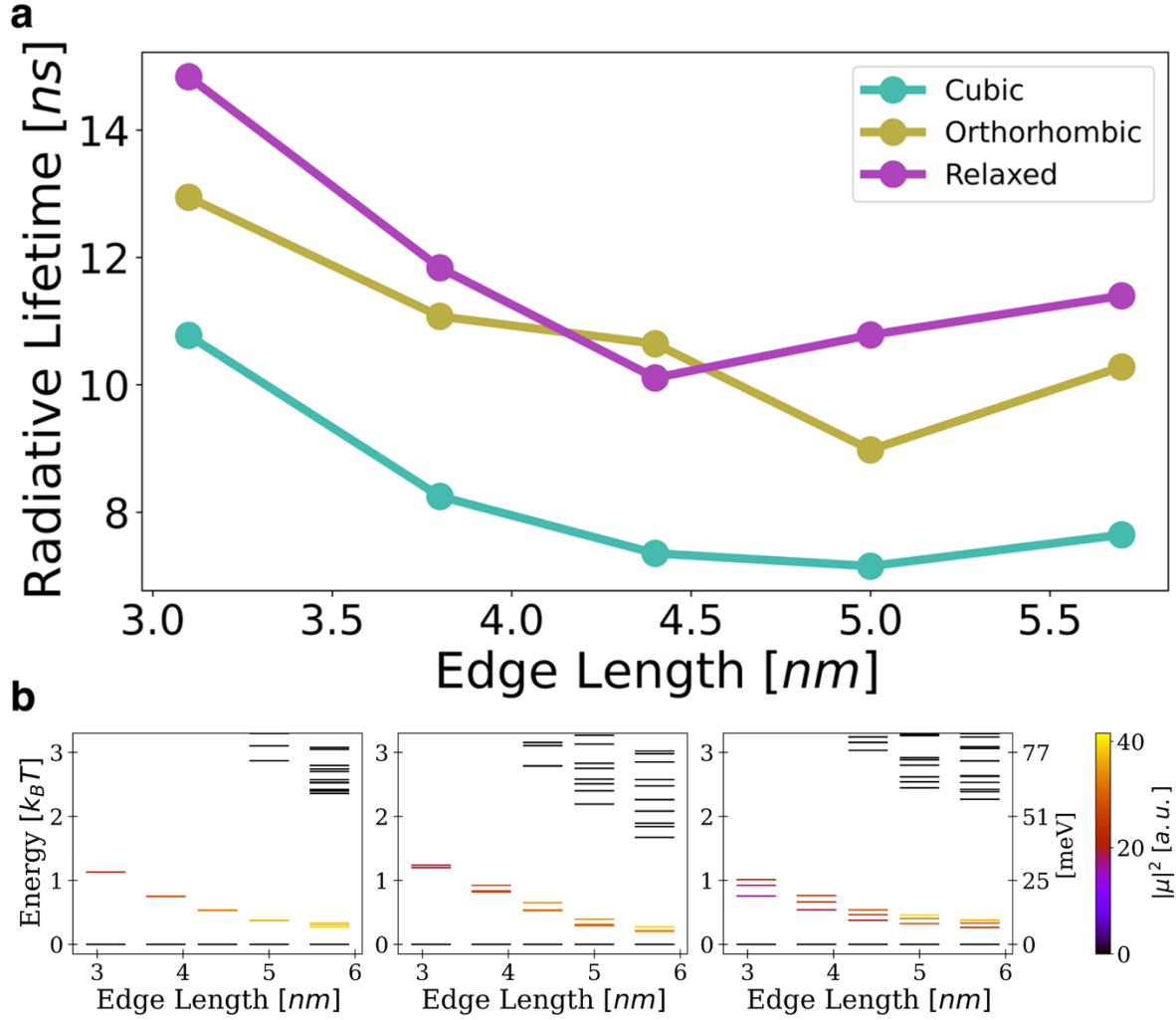

**Figure 5: Size-dependent radiative lifetimes calculation. a**, radiative lifetimes computed through electronic structure calculations on perfect cubic and orthorhombic symmetries along relaxed structure with varying nanocrystal sizes. **b**, Calculated excitonic states as a function of nanocrystal size plotted for cubic, orthorhombic, and relaxed structure. Here, states within $3\kappa_B T$ above the lowest exciton levels are included. Color map indicates oscillator strength, which happens to increase for the lowest bright exciton with increasing nanocrystal size.

# CONCLUSION

In conclusion, the investigation of the effect of $CsPbBr_3$ nanocrystal size on radiative lifetime reveals an unexpected non-monotonic trend. Larger and smaller nanocrystals in the weak and strong quantum confinement regimes displayed longer radiative lifetimes, leaving the intermediately confined nanocrystals to exhibit the shortest radiative lifetimes. This non-monotonicity in radiative lifetime appears to be a common trend among II-VI (CdSe) and perovskites semiconducting nanocrystals. Interestingly, we observed a phase transition from cubic and orthorhombic symmetry that was confirmed by X-ray diffraction and MD simulations. This transition coincided with the intermediate confinement regime, but we find no evidence suggesting its influence on radiative lifetimes. Further analysis revealed that large nanocrystals in the weak confinement regime showed extended radiative lifetime due to the thermal population under ambient conditions of high-lying states with weak oscillator strengths. In contrast, the unexpected lengthening of radiative lifetime in smaller nanocrystals within the strong confined regime was attributed to a significant reduction in oscillator strength, leaving the intermediately confined nanocrystals to exhibit the shortest radiative lifetimes. This study emphasizes the intricate and unique interplay between the $CsPbBr_3$ nanocrystal size and radiative lifetime, paving the way to improve and engineer optical materials with desirable radiative lifetimes for diverse optical applications.

**Data availability**

All the data supporting the findings of this study are available within this article and its Supplementary Information. Any additional information can be requested from corresponding authors.

**Acknowledgements**

This work was supported by Samsung Electronics via Samsung Advanced Institute of Technology (SAIT) under Contract No. FRA000836. Additional support was provided by the U.S. Department of Energy, Office of Science, Office of Basic Energy Sciences, Materials Sciences and Engineering Division, under Contract No. DEAC02-05-CH11231 within the Fundamentals of Semiconductor Nanowire Program (KCPY23). Computational resources were provided in part by the National Energy Research Scientific Computing Center (NERSC), a U.S. Department of Energy Office of Science User Facility operated under contract no. DEAC02- 05CH11231.

**Author contributions**

A.S.A. conceived this study. A.S.A. synthesized and characterized the nanocrystals and conducted steady-state, time-resolved, TEM, and XRD experiments. D.C. developed the order parameter in consultation with D.T.L. D.C. computed the crystal structures using molecular dynamics. D.W. developed the perovskite model for the electronic calculations. D.C. computed the electronic structure and the radiative lifetimes. A.S.A wrote the paper in consultation with all authors. All authors contributed to the revision of the final paper.

**Competing interests**

The authors declare no competing interests.

# Supporting Information

Non-Monotonic Size-Dependent Exciton Radiative Lifetime in CsPbBr$_3$ Nanocrystals

*Abdullah S. Abbas[1,†], Daniel Chabeda[2], Daniel Weinberg[2,3], David T. Limmer[2,3,7], Eran Rabani[2,3,8], A. Paul Alivisatos[1,2,3,4,5,6,§,*]*

[1]Department of Materials Science and Engineering, University of California, Berkeley, Berkeley, California 94720, United States

[2]Department of Chemistry, University of California, Berkeley, Berkeley, California 94720, United States

[3]Materials Sciences Division, Lawrence Berkeley National Laboratory, Berkeley, California 94720, United States

[4]Department of Chemistry, The University of Chicago, Chicago, Illinois 60637, United States

[5]Pritzker School of Molecular Engineering, The University of Chicago, Chicago, Illinois 60637, United States

[6]James Franck Institute, The University of Chicago, Chicago, Illinois 60637, United States

[7]Kavli Energy NanoScience Institute, Berkeley, California 94720, United States

[8]The Raymond and Beverly Sackler Center of Computational Molecular and Materials Science, Tel Aviv University, Tel Aviv 69978, Israel

Present addresses:

[†]Department of Chemistry, The University of Chicago, Chicago, Illinois 60637, United States

[§]Department of Chemistry and Pritzker School of Molecular Engineering, The University of Chicago, Chicago, Illinois 60637, United States

*To whom correspondence may be addressed: paul.alivisatos@uchicago.edu

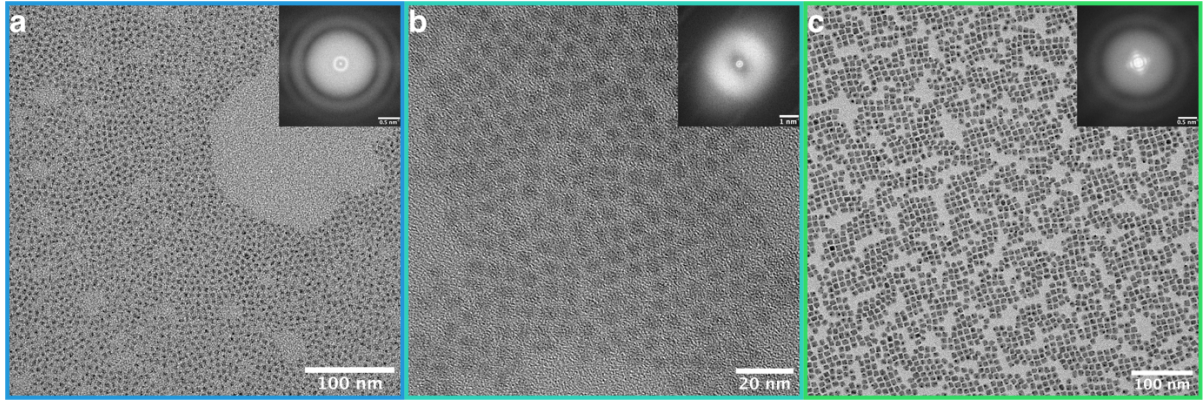

**Figure S1: Transmission electron microscopy (TEM) characterization of CsPbBr$_3$ nanocrystals. a**,**b**,**c**, TEM images of three selected nanocrystals sizes in (**a**) strong, (**b**) intermediate, and (**c**) weak quantum confinement regime with their corresponding Fast Fourier Transform (FFT).

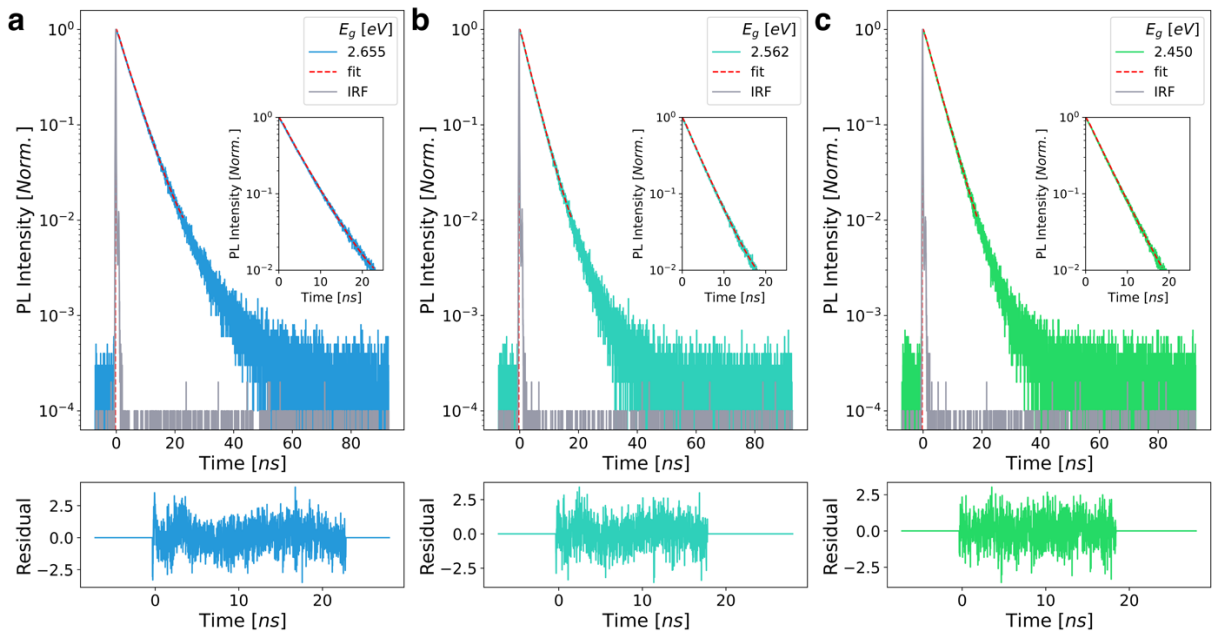

**Figure S2: Time-resolved photoluminescence (TRPL) characterization. a**,**b**,**c**, depict TRPL decays, deconvoluted with instrument response function (IRF), with the initial two decades fitted to a monoexponenital function for nanocrystal sizes within (**a**) strong, (**b**) intermediate, and (**c**)

weak quantum confinement regime. Each plot includes the residuals showing the goodness of the fit.

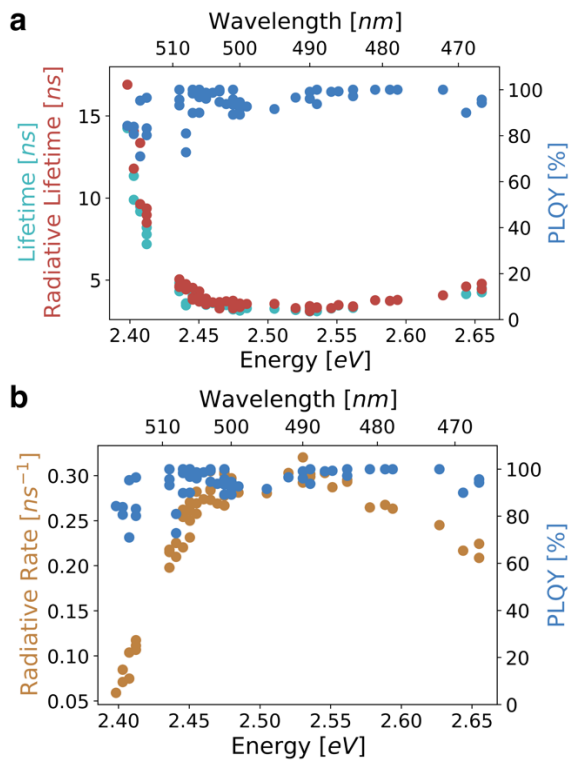

**Figure S3: Size-dependent radiative lifetimes and rates of CsPbBr$_3$ nanocrystals. a,b**, Varying nanocrystals spanning further in the weak confinement regime illustrating extended (**a**) radiative lifetimes and reduced (**b**) radiative rates.

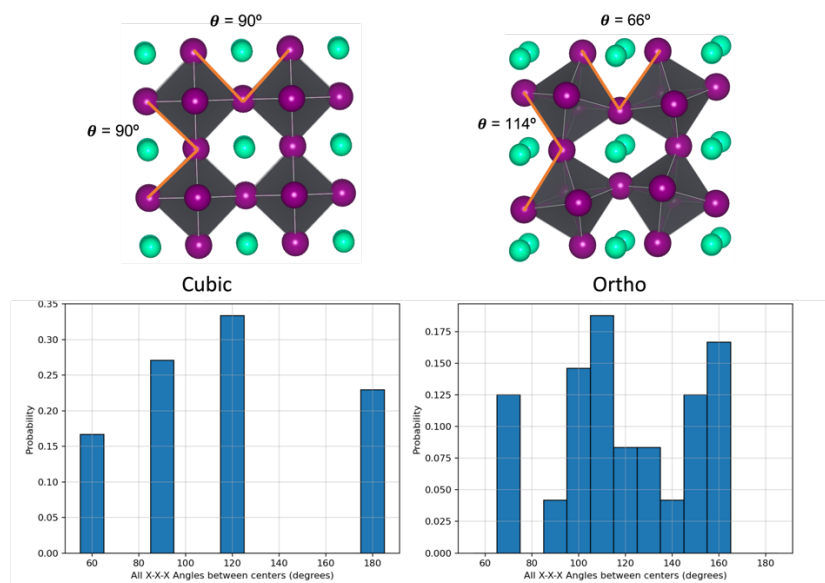

**Figure S4: Cubic vs orthorhombic CsPbBr$_3$ structure. a,b**, Probability distribution of X-X-X angles in ideal (**a**) cubic and (**b**) orthorhombic CsPbBr$_3$ structure. In the cubic symmetry, the 60° angle is distinctly prominent, while 160° angle occurs only in the orthorhombic symmetry.

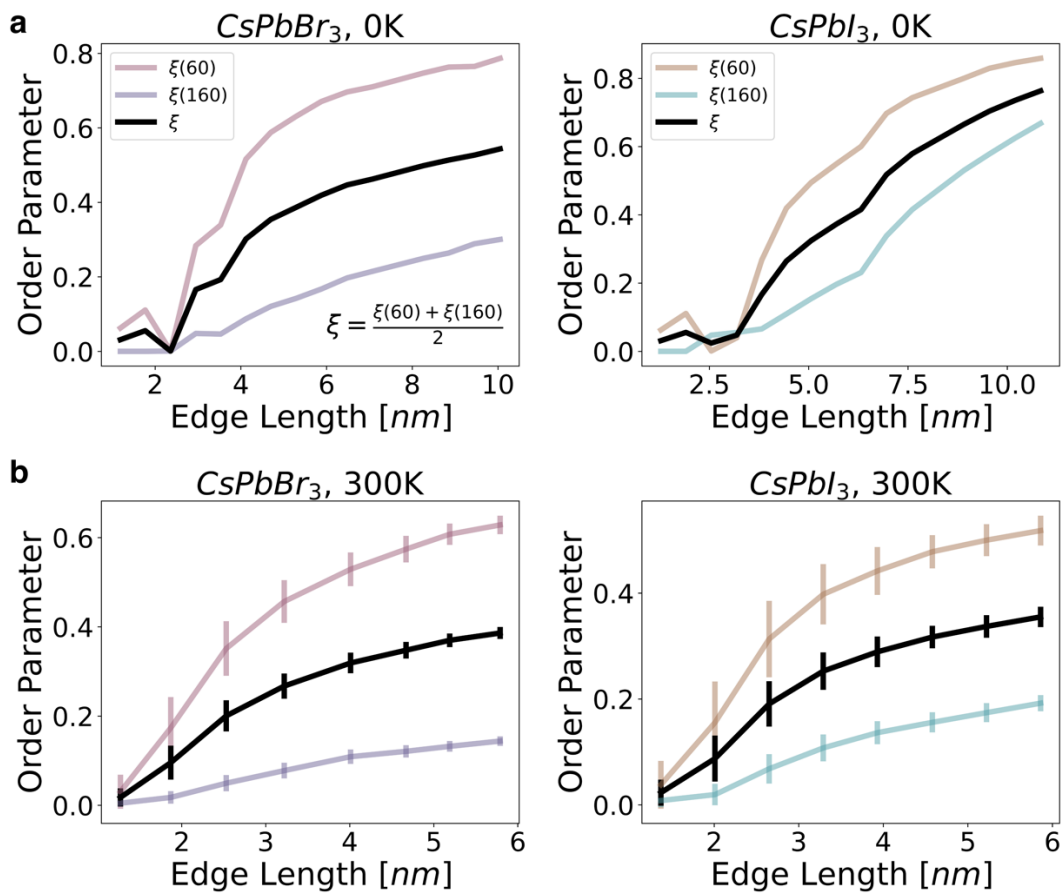

**Figure S5: Order parameter. a,b**, Comparison of order parameter for $CsPbBr_3$ and $CsPbI_3$ nanocrystals with varying sizes at (**a**) 0 K and (**b**) 300 K. Both perovskites initially exhibit cubic symmetry in smaller sizes, then transition into orthorhombic symmetry as the nanocrystal size increases.

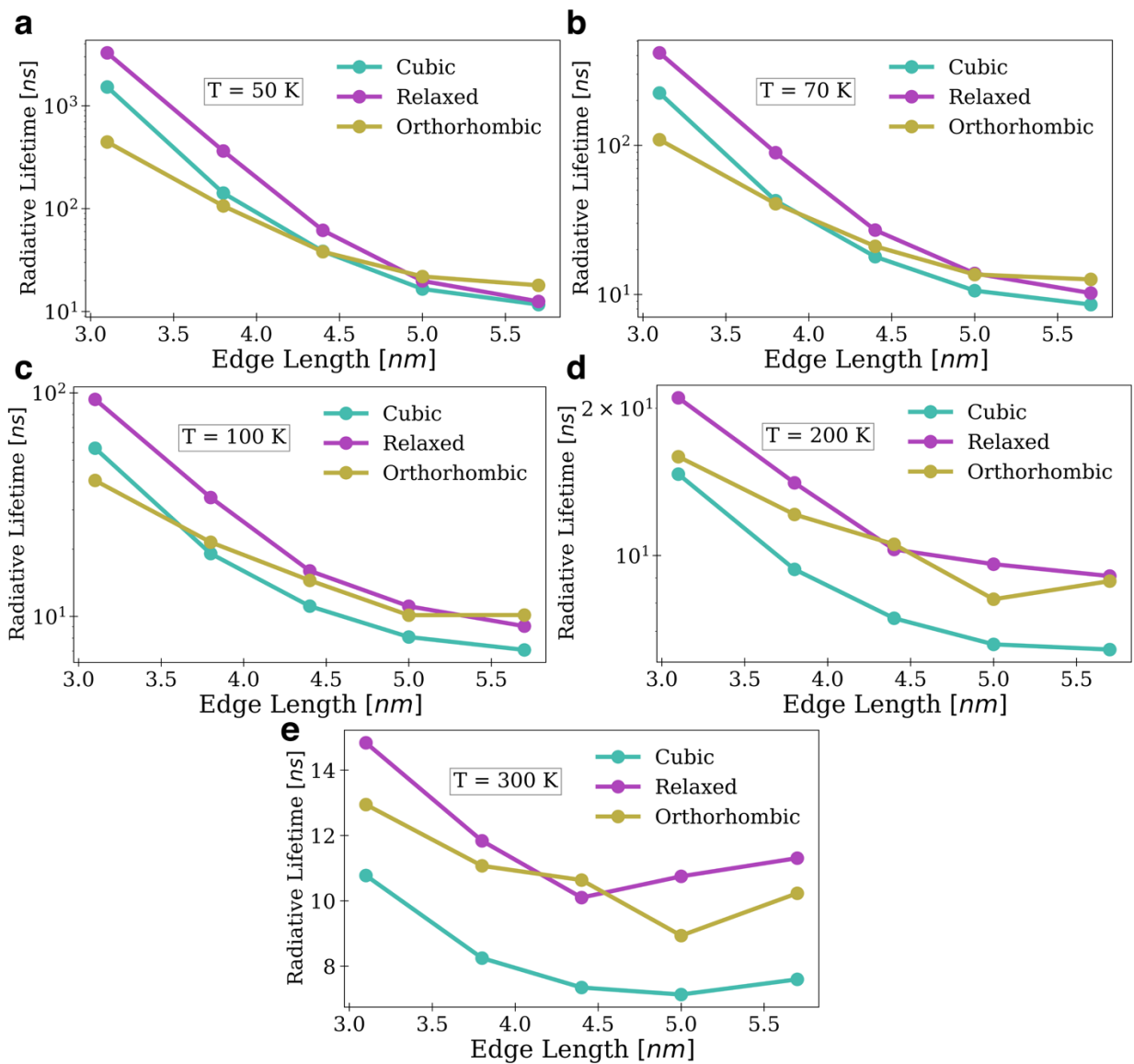

**Figure S6: Size Dependent Radiative Lifetime Across Different Temperatures. a,b,c,d,e,** Radiative lifetimes of halide perovskites as a function of temperature calculated at (a) 50 K, (b) 70 K, (c) 100 K, (d) 200, and (e) 300 K, demonstrating a transition from an exponential trend observed for low temperatures to a non-monotonic trend for higher temperatures, indicating that the dim states became thermally accessible.

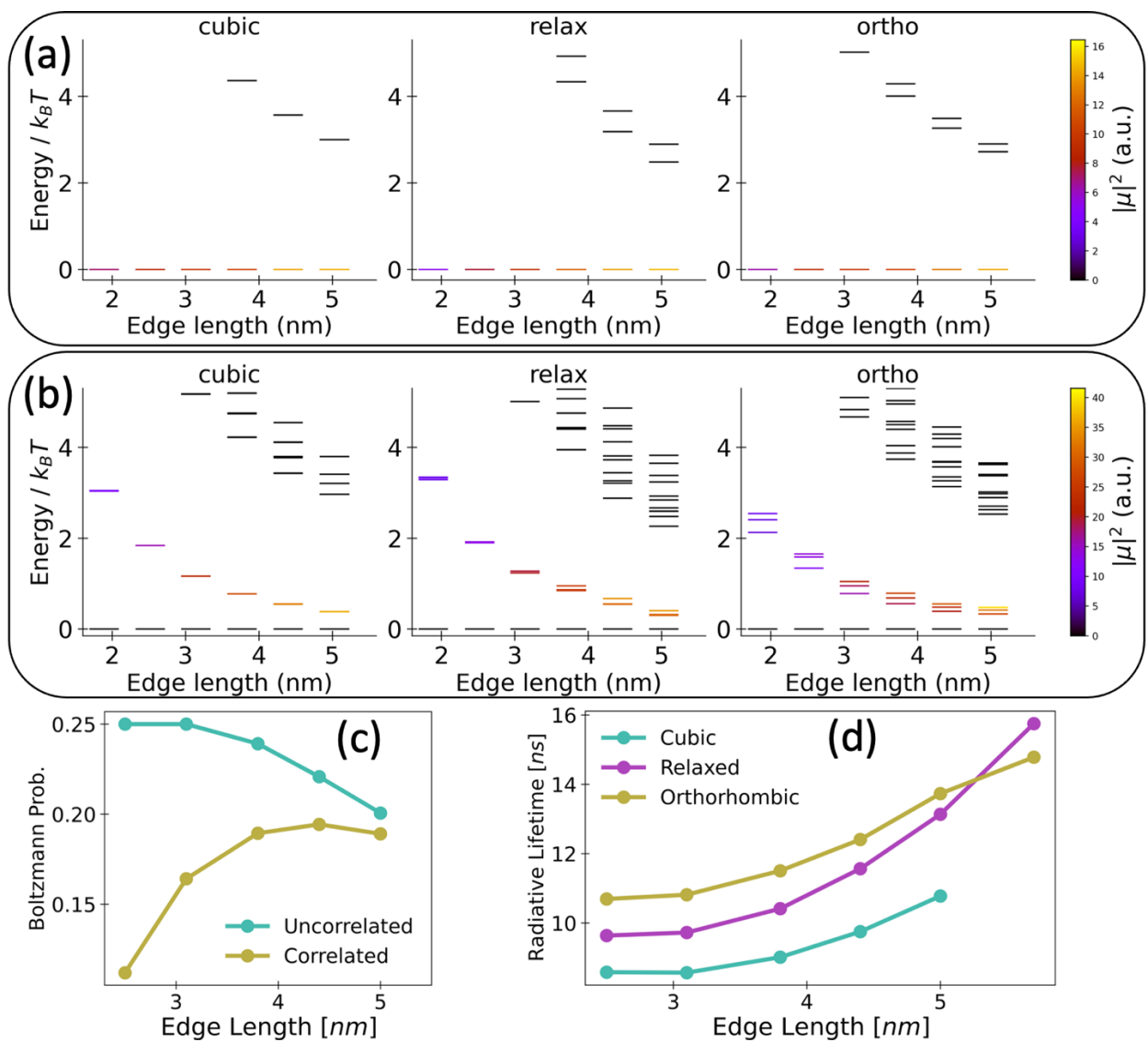

**Figure S7: Contrasting Size Dependent Radiative Lifetime Trend Without Electron-Hole Correlation.** a,b,c,d, Exciton energy level diagrams as a function of CsPbI$_3$ nanocube edge length for (a) uncorrelated and (b) correlated excitons. For uncorrelated electron-hole pairs, the lowest energy states are quadruply degenerate bright states, with higher lying dim states becoming thermally accessible for larger systems in which the DOS is increased. The fine structure changes when electron exchange is included: the lowest energy state is a dim singlet below a manifold of bright triplet excitons. (c) Size-dependent Boltzmann population of the lowest energy bright state for uncorrelated (blue) and correlated (gold) excitons. For small NCs, there is opposing size

dependent behavior; thermal population of higher lying states decreases the emissive ground state population for uncorrelated excitons, while for correlated excitons the emissive triplet state increases in population for larger NCs, until the inflection point where thermal population of higher lying dim states decreases the triplet population. (d) Predicted radiative lifetimes of uncorrelated excitons increase monotonically due to decreased Boltzmann population of the emissive manifold. Contrastingly, for correlated excitons, reduced confinement in larger NCs lowers the energy of the bright triplet states with respect to the dim ground state, increasing the Boltzmann population of emissive states and decreasing the lifetime (see Fig. 5).